\documentclass[12pt]{article}
\usepackage{amsfonts}
\usepackage{amssymb}
\usepackage{amsmath}
\usepackage{textcomp}
\everymath{\displaystyle}
\usepackage{cancel}
\usepackage[pdftex]{graphicx}
\usepackage{graphicx}
\usepackage{graphics}
\usepackage{tikz}
\usepackage{titlepic}
\usepackage{pgfplots}
\pgfplotsset{width=9cm,compat=1.9}
\usepackage{enumitem}
\usetikzlibrary{intersections}
\usetikzlibrary{
  knots,
  hobby,
  decorations.pathreplacing,
  shapes.geometric,
  calc
}
\usepackage{braids}
\usepackage{multicol}
\usepackage{cancel}
\usepackage{longtable}
\usepackage{subfigure}
\usepackage{indentfirst}
\usepackage[utf8]{inputenc}
\usepackage{tikz-feynman}

\def\be{\begin{equation}}
\def\ee{\end{equation}}
\def\beq{\begin{eqnarray}}
\def\eeq{\end{eqnarray}}
\def\bn{\begin{eqnarray*}}
\def\en{\end{eqnarray*}}

\def\P{\Phi}
\def\p{\phi}

\def\O{{\cal{O}}}

\def\pd{\partial}

\def\m{\mu}

\def\l{\lambda}

\def\cL{{\cal{L}}}

\def\cO{{\cal{O}}}

\def\cS{{\cal{S}}}

\title{Supermassive blackhole-Ultralight dark matter connection 
Essay written for the GRF 2024 essay competition
}

\author{T R Govindarajan\footnote{trg@imsc.res.in}\\
The Institute of  Mathematical Sciences, Chennai, India\\
Krea University, Sricity, India\\
{\textcolor{red}{\em selected for honorable mention}}}

\date{\today}
\begin{document}

\maketitle

\begin{abstract}
A novel proposal linking the origin of 
supermassive blackhole formation and ultralight dark matter. 
Ultralight bosons of mass $10^{-22}-10^{-24}$ eV 
which is one of the front running proposal for dark matter could 
also linked with the formation and evolution of supermassive 
blackholes at the center of all the galaxies.
\end{abstract}

\section{Introduction}
It is established that galaxies are held together by mostly
non-luminous matter known as dark matter\cite{darkmatter}. 
The radial speeds of 
of stellar objects in our own and other galaxies do not 
drop down fast enough as expected in Newtonian gravity led to the 
speculation of such matter. In addition velocity dipersions,
galaxy clusters, gravitational lensing,
bullet clusters and CMB offer further support for such matter. 
The important constraint on the candidate dark matter is,  it should 
be only coupled extremely weakly to the normal matter and 
gravity is the only link between the two sectors. There were 
several proposals from  High energy physics beyond 
the standard model. Supersymmetric extensions provide several 
potential massive particles to fill this gap. Unfortunately 
they do not provide proper densities in the expanding universe
evolution from big bang. There are proposals for sterile neutrino
or QCD axion to explain this puzzle and they also do not 
match correctly. New dark photons were also suggested and do not 
have experimental signals too. Primordial blackholes  also have been 
proposed and fail to have any convincing follow up\cite{dmsummary}. 

It is also suggested to modify gravity at extremely low accelerations
away from Newtonian gravity which is seriously followed\cite{mond}. 
But such modifications do not stem from some fundamental principles
and also have unanswered questions. We will comment on these 
later in our conclusions.

A resolution on this question begs to look for some fascinating 
inputs from  quantum theory in combination with gravity, 
though may not be 
directly connected to difficult questions of quantum gravity.

The suggestion is to introduce untralight bose particles 
whose mass is $\leq \O(10^{-22})eV$ which could provide the 
necessary wave like background and explain several features. 
Being ultralight, it is necessarily relativistic in nature and 
is naturally expected to escape at the early stages of evolution
to be of any use. But quantum theory of bose gas comes to the 
rescue through the critical temperature at which Bose Einstein
condensate could be formed and such collections of  
lumpy quantum objects offer a novel solution to the dark matter 
question\cite{ultralight}. 
Naturally it should be chargeless to escape being 
detected through electromagnetic interactions. We will provide a 
natural candidate which has  its origin to the queries of 
Schrodinger himself\cite{schrodinger}.

In this background  we also have yet another unexplained  question
related to the formation of galaxies very early after the bigbang. 
Now there are evidences early stars and galaxies were there 
around 300-400 million years after the bang\cite{JWST}. 
These galaxies do also 
possess central supermassive blackholes at the center and their 
formation is also expected to be 
tied up with dark matter. Our proposal naturally links 
these two unanswered questions and provide possible 
resolution. Word of caution: More work needs to be done to 
establish to build the scenario. 

In section 2 we will provide our proposal for dark matter question
and build evidences towards that in section 3. Section 4 will 
discuss about the seeds of primordial blackholes to grow along 
at the early times itself. Section 5 will provide a discussion 
of the strengths and weaknesses of the proposal and conclude with  
future directions.
\section{Proposal for dark matter} The ultralight dark matter
proposal needs a scalar boson which does not interact with matter 
except through gravity. The anticipated mass is less than 
$\O(10^{-22})eV$. The only particles which we know obey such 
bound is photon and graviton itself. Both are expected to be 
massless within the standard model. A question was posed by 
Schrodinger in 1955, `Should the photon be massless?'
\cite{schrodinger} 
He answered that it need not be,  provided its coupling to matter 
is through a conserved current. That is if $A_\mu$ is the vector 
potential  it should be coupled to the current of the matter $j_\mu$ 
through the interaction term $e~j_\mu~A^\mu$ where $\pd_\mu~j^\mu~
=~0$. He also provided a mass bound for photon to be $\leq 10^{-16}
eV$. He used the data from geomagnetic fields which gives 
a natural length scale of the radius of the earth which serves 
as the scale of the Compton wavelength. This has been improved 
with better geomagnetic data to $\leq 10^{-18}eV$. Incidentally 
this is the bound given in Particle Data Book. Using 
galactic magnetic fields there are limits 
$\cO(10^{-25})eV$ which are yet to be verified.
But there are couple of problems  with having mass for the photon
however small it is.. 
First is gauge invariance is lost. Second is that there is a 
discontiuity in the degrees of freedom from 3 to 2 in the massless
limit. This was resolved by Stueckelberg earlier by 
introducing another scalar boson. This can be thought of as 
precursor to mass through sponteneous symmetry breaking.      
A simple action  was provided by Stueckelberg\cite{stuckelberg}.
\be
\cS~=~\int -\frac{1}{4}F_{\m\nu}~F^{\m\nu}~+~\frac{1}{2}~m^2~
\left(A_\m~-~\frac{1}{m}\pd_\m\p\right)^2 
\ee
As pointed out by Schrodinger it can be coupled to matter through 
a conserved 
current. This theory of QED will not contradict with any of the 
results if $m~\leq 10^{-18}eV$ at the current levels of 
measurements. In addition it will indeed help in understanding 
infrared problems of QED. Subtle massless limit of such a theory will 
naturally provide the Faddeev-Kulish realisation of QED\cite{trg}. 
One can understand this through Inonu-Wigner contraction 
of the little 
group of massive spin -1 particle, namely $SO(3)$ going over to 
Eucliden group $E(2)$ the little group of massless representation.
\subsection{The dark matter question} Having given the background 
we have for the massive spin 1 photon three degrees of freedom, 
an extra longitudinal component which using gauge invariance can 
be decoupled to the Stuckelberg scalar field. This field is proposed 
as candidate for ultralight dark matter. It couples to the 
matter fields only extremely weakly with wavelengths of the 
$\cO\left(\frac{h}{mc}\right)$, the Compton wavelength. 
This ultralight 
bosons can condense to form BEC condenstae below the  
crtical temperature $T_c$. 
\subsection{Critical temperature $T_c$}      
We have to consider relativistic ideal gas for estimating the 
critical temperature \cite{Tc}. It is given by:
\be
\label{TC}
T_c =  \frac{\hbar c}{k_B}
\left(\frac{\rho\pi^2}{m_{\gamma}\zeta(3)}\right)^{1/3}.
\ee
Given the mass of the photon to be $\leq 10^{-18}eV$ we parametrize 
the mass as $10^{-22+x}eV$. Given the current estimate of the 
dark matter energy density as $\cO(1)$ 
Gev/cc the $T_c$ can be expected 
to be $10^{17}-10^{19} K$. This is close to the temperature 
at the big bang and the condensate of Stuckelberg photons 
will be formed 
at the earliest times itself. The time could be estimated  to 
be Lepton era at around $10^{-15} sec$. This radiation dominated time
scale makes the transverse photons equilibrium by about 50K years. 
But the longitudinal/Stuckelberg  photons will 
spread to the edge of the 
universe at that time and condense. 

Given this background it is clear at the very early universe ultra 
light dark matter condensate provides the much needed pressure to 
assist the formation of galactic structure even at the early times.
But still the core of the galaxies namely the supermassive blackhole 
should be formed without the need for nuclear fusion and stellar 
evolution pathways. This should happen only by the gravity and it 
needs the seeds of such blackholes. Towards this we enhance 
Stueckelberg theory to Abelian Higgs mechanism in sec 4.
\section{Support for Ultralight scalar}
We will provide some evidences in favor of ultralight scalar.
Ultralight scalar as Stuckelberg particle with mass in the 
range of $10^{-18}-10^{-24}eV$ seems to be favored \cite{hui}. 
\begin{enumerate}
\item 
This corresponds to Compton length of $3\times 10^3$ 
light seconds to $3\times 10^9$ light seconds which is 
$0.0001~-~100$ light years. The size of the 
BEC condensate is expected to be close to that of compton wavelength.
If that were the case the size of dwarf galaxies should be greater 
than this size. There are expected 200 dwarf galaxies around 
the Milky way (through N-body simulation). There are around 50 
galaxies seen with various sizes ranging from 500 light years 
to 1,00,000 light years holding around 1000 - 1000000  stars. The 
not seen or missing one may be  smaller in sizes with very low 
luminosity. The smallest galaxy seen so far is Segue 2 
with half radius of 115 light years.
Our expectations predict the lower bound on the size to be 100 
light years. 
\item The ultra light scalar also provides profile of halos
of dark matter which does not have the cusp problem at the core.
Following the work of Hui etal., \cite{hui} 
we can expect the mass for 
ultralight scalar can fit with a range of $10^{-22}-10^{-24}eV$.
\item Current pulsar timing array observations indicate a background
noise due to early merger of galaxies. These affect the pulsars 
in our galaxy as expected from Rubakov \cite{rubakov}. 
FDM affects the time of 
arrival at different pulsars. This also provides measure 
of dark matter density. The frequencies and amplitudes provide 
an upper bound on the FDM masses to be $\leq 10^{-24}eV$. It would be 
nice by precision measurements to reduce the uncertain width.
Note if photon masss is the factor for the dark matter, then this 
provides an upper bound for it. 
\item Ryutov etal\cite{ryu} considered the 
extra pressure due to longitudinal 
photons of massive electromagnetic theory. For correlations 
of length scales compared to the Compton wavelength when longitudinal
photons affect the pressure, they find the extra pressure mimicking 
the dark matter. 
\end{enumerate}
\section{Abelian Higgs mechanism}
Stuckelberg mechanism for providing mass for the photons without
breaking gauge invariance can be obtained as a limit of
the gauge field coupled to a complex scalar field with
spontaneous symmetry breaking. Consider $\Phi$ a complex scalar
field and U(1) gauge theory.
\bn
\cL~&=&~-\frac{1}{4}F_{\m\nu} F^{\m\nu}~+~|D_\m~\Phi|^2~-~V(\Phi)\\
V(\Phi)~&=&~-\m^2|\Phi|^2~+~\l~|\Phi|^4
\en
We transform to polar coordinates for the complex scalar field,
$\P~=~R~e^{i\p/v}$ and we get
\be
\cL~=~-\frac{1}{4}F_{m\nu}^2~+~
m^2\left(A_\m~-~\frac{1}{m}\pd_\m~\p\right)^2~+~
\frac{1}{2}(\pd_\m~R)^2~-~\m^2~R^2~+~f(\{R,\p\})
\ee
If $R$ is frozen at high mass value, $R^2~=~\frac{2v\l^2}{2}$
where $\l$ is the quartic coupling constant and v is the vacuum
expectation value $\left(v^2~=~\frac{\m^2}{2\l}\right)$ we  get the
mass of the photon to be
$m~=~ev$. Stuckelberg theory is the limit of Higgs mass $\rightarrow
\infty$. In principle we can set the Planck mass as this limit.
So in addition to massive photon of extremely small mass
$\leq 10^{-18}eV$ we will have extremely heavy new scalar Higgs.
This is the scenario at the early universe when ultralight scalars
are formed. The Higgs will act as seeds for the primordial
blackholes which provide center of attraction for galaxy formation.
\subsection{Supermassive blackholes and dark matter}
Having provided the role of massive photon mass FDM we  
now look at the Higgs part of the complex scalar field 
whose mass is very large as the seeds of the primordial blackholes
\cite{pbh} 
simultaneously with dark matter. The formation galaxies earlier 
than the initial expectations with 
supermassive blackholes at the core 
provide support for the speculations that these could be the cause of 
such objects. The strength of such queries can be answered through
dwarf galaxies particularly those with SMBH at the core and 
star formations still taking  place.    
 
We now have primafacie evidence for SMBH at the center of ultra 
compact dwarf galaxies in Virgo clusters. The sizes are $\cO(100)$
parsecs. The SMBHs have masses $10^6$ solar masses. They also 
contain large proportion of the galactic masses to the 
extent of $\geq 10\%$. 

In recent JWST observations at the earliest times 
of 300-500 million years after the Big bang have led to the 
astonishing formation of baby galaxies \cite{jwst}. 
These could not have been possible 
without the seeds being born very close to the times of bang itself.  
\section{Conclusions}
We have argued in favor of the Stuckelberg field which provides 
the mass of photons could be candidate for ultralight matter.
It could also be considered as axion like dark matter field.
The highly massive Higgs field produced along with that close to the 
big bang promises to provide the seeds primordial blackhole 
which could be the origin of SMBH at the early stages of the 
universe. This proposal has interesting prediction about the size 
of the dwarf galaxies as well as signals from pulsar timing array. 
Detailed developement of the proposal could provide new features 
in CMB and Nanogravitational waves. Lastly the effective field 
approach of the background of the dark matter profile background 
can be the source of the deviation needed for MOND.

\end{document}